# Newtonian Relativity, Gravity, and Cosmology

Joseph L. McCauley+
Department of Physics
University of Oslo
Box 1048, Blindern
N-0314 Oslo

and

Institute for Energy Technology
Box 40
N-2007 Kjeller
jmccauley@uh.edu

## Abstract

Well-known to specialists but little-known to the wider audience is that Newtonian gravity can be understood as geodesic motion in space-time, where time is absolute and space is Euclidean. Newtonian cosmology formulated by Heckmann agrees implicitly with Cartan's formulation but does not unfold the underlying geometric picture in space-time. I present the transformation theory of Newtonian mechanics and gravity developed by Cartan and Heckmann, and show via coordinate transformations that Heckmann's Newtonian cosmological model has a center, so that the cosmological principle cannot hold globally.

It is possible to confuse relativity principles with a position of relativism. Mach's principle has much to do with the latter and little to do with the former. Both general relativity and nonlinear dynamics inform us that most coordinate systems are defined locally by differential equations that are globally nonintegrable: global extensions of local coordinates usually do not exist. I explain how defining inertial frames locally by free fall short circuits Mach's philosophic objections to Newtonian dynamics.

[+] permanent address:  Physics Department
                        University of Houston
                         Houston, Texas 77204 USA

## 1. Newtonian dust

In Newtonian theory we can consider a pressureless dust obeying the coupled quasilinear equations of hydrodynamics

$$\frac{d\vec{v}}{dt} = \frac{\partial \vec{v}}{\partial t} + \vec{v} \cdot \nabla \vec{v} = -\frac{1}{\rho} \nabla \Phi$$

$$\frac{\partial \rho}{\partial t} + \nabla \cdot \rho \vec{v} = 0$$

$$\nabla^2 \Phi = 4\pi\rho \qquad . \quad (1)$$

In a cosmological model each dust particle represents a galaxy. The connection with Newtonian particle dynamics is provided by the method of solution of (1), the method of characteristics. Equations (1) are quasilinear partial differential equations whose characteristic curves [1,2] are generated locally by the differential equations

$$\frac{dt}{1} = \frac{dx_k}{v_k} = \frac{dv_k}{-\frac{\partial \Phi}{\partial x_k}} \quad . \quad (4)$$

These are simply the equations of Newtonian mechanics: if we think of N dust particles then each dust particle obeys Newton's law in the gravitational field defined by the other N-1 dust particles. We can rewrite the characteristic equations in the form

$$\dot{x}_k = v_k$$

$$\dot{v}_k = -\frac{\partial \Phi}{\partial x_k} \qquad (4b)$$

so that we might attempt to study the nonlinear dynamics of galaxy formation and evolution in the 3N dimensional phase space of the x's and v's, or as Newton's second law

$$\frac{d^2 x_i}{dt^2} + \frac{\partial \Phi}{\partial x_i} = 0 \qquad (5)$$

in 3-space. We can think of streamlines in a 6-dimensional (x,v) phase space (phase flow picture) if and only if the solutions x and v are finite for all real finite times, in which case all singularities of power series solutions of (5) are confined to the complex time plane. We will see below that this condition is violated by spontaneous singularities in Newtonian cosmology. In this case the language of jet space and caustics [3] provides the right approach to the analysis of the nonlinear dynamics of galaxy motions.

Newtonian cosmology is still of interest for at least two main reasons. First, it reproduces the same equation for the Hubble expansion as the Friedmann model in general relativity. Second, and more fundamentally, Cartan's formulation and interpretation of Newtonian dynamics provides the best take-off point for general relativity. The problem of defining average solutions in theoretical cosmology is unsolved in general relativity although much headway has been made in Newtonian cosmology. It is hoped that perturbation theory in Cartan's formulation of Newtonian theory may shed some light on how to handle the problem in general relativity [4].

## 2. The cosmologic principle

The Hubble expansion is inferred nonuniquely from redshift data. The galaxy distribution is obtained from the analysis of data derived from redshift data. The analogy of galaxies in an expanding universe with points on an expanding balloon is well-known. In the Hubble's law interpretation of redshifts the matter distribution of the observable part of the universe is implicitly presumed to be isotropic and homogeneous, so that galaxies appear to recede from one another with a steady radial velocity field [5]

$$v = H(t)r \qquad (6)$$

where $H(t)$ is the very slowly-varying Hubble 'constant' and is given approximately by $H(t) \approx h^{-1} 100 km/s.Mpc$, with $h \approx .5$ to $.6$ in our present epoch. Since $H(t)$ varies with time the presumed equivalent observers are accelerated relative to one another. The main point to bear in mind is that this simplest form (6) of the Hubble law implicitly presumes the cosmological principle, so that the usual method of data analysis in cosmology would have to be revised or abandoned if the cosmological principle could be shown to be false. For example, one does not yet know how to generalize (6) to handle the case of a hierarchical universe, or any other nonhomogeneous universe.

The cosmologic principle is sometimes stated in the following form: All observers are equivalent; there are no preferred observers: there is no preferred vantage point from which to observe the gross motions of matter in the universe. This statement is not precise. I will emphasize in part 5 below that the standard definition of equivalent observers simultaneously defines those observers as *preferred*: from their globally accelerated (but locally inertial) reference frames, and from no others, the Hubble expansion would look the same.

Another way to express the content of the cosmological principle is to assume that the distribution of matter in the universe is globally (more or less) homogeneous and isotropic. Equivalently, one can use the assumption that the density is spatially constant at fixed times, with only small deviations from uniformity.

The cosmological principle appears superficially to resemble a principle of relativity, but that principle is neither the basis for the general theory of relativity nor is it demanded by that theory. The cosmological principle is not required by any known law of physics. Some writers allude to 'the Copernican Principle' but that principle is merely another statement of belief in the cosmological principle (Copernicus certainly did not adhere to the 'principle' attributed to him, because his own model universe had a center). I defer the discussion of the observational basis for or against the cosmological principle until a second paper, and discuss here only the attempt to realize the cosmological principle within an infinite Newtonian universe.

Newtonian cosmologies require flat space (a flat space is one where Cartesian coordinates exist globally because the Riemann curvature vanishes everywhere). Flat spaces (and Newtonian mechanics) can be realized in two ways: (i) as an infinite unbounded space, or (ii) as a finite unbounded space in the form of a flat 3-torus. The latter is equivalent to solving (1) with periodic boundary conditions, a common procedure in numerical simulations of the N-body problem.

I review next one of the main building blocks of Newtonian (and Einsteinean) cosmology: the principle of equivalence. The discussion leads us into the transformation theory of Newtonian mechanics, and to Cartan's formulation of Newtonian theory, which reflects the correct geometric nature of gravity: Cartan's formulation agrees with the infinite light-speed limit of general relativity, whereas the standard textbook interpretation of Newtonian gravity as a scalar potential giving rise to a covariant force does not. In order to arrive well-informed at the goal it's necessary to start at the beginning.

**3. Free fall and the principle of equivalence**

Consider N particles of fixed mass $m_i$ interacting only via gravity in what follows. For any one of the N bodies the law of inertia states that

$$\frac{dv_i}{dt} = 0 \quad (7)$$

whenever no net force acts on the body, and in a gravitational field with potential $\Phi$ Newton's second law

$$\frac{d^2 x_i}{dt^2} + \frac{\partial \Phi}{\partial x_i} = 0 \quad (5)$$

generalizes the law of inertia. I assume in all that follows that the potential $\Phi$ is independent of the mass $m_i$ of the body described by (5).

Standard treatments of Newtonian mechanics assume, without explanation, that the gravitational potential transforms like a scalar (Newton did not tell us how gravity should transform under arbitrary coordinate changes). The gravitational force is then forced to transform like a covariant vector. This assumption is not only unnecessary, it also does not agree with the infinite light-speed limit of general relativity. When it is used then the transformation theory of Newtonian mechanics does not emphasize the following fundamental fact about gravity: a reference frame in free-fall is locally inertial. In a freely falling frame a ball tossed into a vacuum obeys the law of inertia locally (not globally), is force-free and has a trajectory in the freely falling frame that is a straight line with constant speed (by local I mean the consideration of a small enough spatial region that the gradients of the gravitational field can be neglected for a short but finite time, whereas global refers to arbitrarily large spatial regions and time intervals). The experiment could be performed by an astronaut near a satellite falling freely about about the earth, or by a robot made to jump off a cliff on Mars. Treating the gravitational force as a covariant vector emphasizes that gravity cannot be transformed away globally (tidal forces cannot be transformed away, e.g.). We want to use instead a transformation theory that explicitly makes use of the fact that the gravitational force can be made to vanish locally.

The second related fact, emphasized by Einstein, is the equivalence principle. If we transform from a local inertial frame (meaning here any freely falling frame, or any frame connected by a Galilean transformation to a freely falling frame) to a linearly accelerated one,

$x'_i = x_i - b_i(t)$

$v'_i = v_i - \dot{b}_i(t)$

$a'_i = a_i - \ddot{b}_i(t)$,     (8)

where $db_i/dt$ is constant for Galilean transformations, then Newton's second law becomes

$$\frac{d^2 x_i}{dt^2} + \ddot{b}_i + \frac{\partial \Phi}{\partial x_i} = 0.$$    (5b)

When $d^2 b_i/dt^2$ is constant then this is the same as if the accelerated frame were replaced by a frame that is stationary in a gravitational field with local field strength $g_i = d^2 b_i/dt^2$, corresponding to a planet whose surface is transverse to the $x_i$-axis. By giving up the fiction that the gravitational force transforms globally like a covariant vector we shall see that both of these fundamental empirical facts about how gravity behaves *locally* can be adequately emphasized within the transformation theory of Newtonian dynamics.

**4. Newtonian gravity transforming as a scalar potential**

Given Newton's second law, if one asks for the force that produces an ellipse with the point of attraction at one focus, then the result is the inverse square law of attraction. Because this method of derivation of gravity is due to Newton, he surely knew that his description of gravity as an inverse square law force is accurate only to the extent that the orbits of planets can be accurately described as (nonprecessing) ellipses [6]. Note that no assumption was made here about how gravity behaves under arbitrary coordinate transformations. I next review the usual (and wrong) treatment of the transformation theory of Newtonian dynamics, where gravity is supposed to transform as a scalar potential.

Assume that Newtonian gravity can be described under arbitrary coordinate transformations as a scalar potential $\Phi$. This means that under differentiable coordinate transformations

$$x'_k = f_k(x,t), \quad x_i = g_i(x',t) \qquad (9)$$

the gravitational potential transforms like

$$\Phi'(x',t) = \Phi(x,t) \qquad (10)$$

which seems obvious from a superficial standpoint[1]. The gravitational force $\vec{F} = -\nabla\Phi$ must then transform like a covariant vector, because

$$\frac{\partial \Phi'}{\partial x'_k} = \frac{\partial x_i}{\partial x'_k}\frac{\partial \Phi}{\partial x_i}. \qquad (11)$$

It follows that if $\vec{F} = -\nabla\Phi = 0$ globally in one frame then $\vec{F}' = -\nabla'\Phi' = 0$ globally in all (global) frames. This global perspective is misleading because it masks the all-important fact that the gravitational force vanishes in local inertial frames that are defined by free fall. Why does this matter?

---

[1] A scalar that is, in addition, invariant would transform like $\Phi(x',t) = \Phi(x,t)$.

There are no known global inertial frames (stars and galaxies aren't fixed, but accelerate). All known inertial frames are local, and the only local inertial frames that we know in nature are those provided by free fall. By local, I mean simply that the gravitational field is approximated over a small but finite region by a constant, so that the derivatives of the field are ignored (tidal effects are global and are therefore ignored). The earth, in free fall about the sun, is a good local inertial frame over small spatial regions and over times that are short compared with the rotation period of the earth about it's axis. Were this not true then Galileo could not have discovered the local laws of free fall and inertia from his borderline medieval perspective of Archimedean empiricism combined with neo-Platonic argumentation. Galileo understood the law of inertia as a local, not global principle, but for the wrong reason [7]: he did not deviate from Plato and Aristotle in regarding uniform circular motion as 'natural', and as requiring no mechanical explanation (Aristotle and Galileo also required no explanation for gravitational free-fall and radial coalescence of mass via gravity). Galileo largely regarded constant speed linear motion (force-free motion, which we identify as the law of inertia) as only a local tangential description of a global orbit that he believed (in agreement with Plato and Aristotle) should be circular, at uniform speed. Galileo described parabolic trajectories of canon balls but did not realize that gravity also makes the planets accelerate in approximately circular orbits about the sun. Newton was the first to make the connection between the trajectories of apples and the orbit of the moon, and beyond, although Huyghens preceded Newton in using the law of inertia (for tangential motions) combined with the second law (for radial motions), before Newton, to prove that an inverse square force of attraction is necessary for uniform circular motion. Galileo's local dynamics was new and anti-Aristotelian, but was inconsistent with his global view of astronomy, which remained Copernican-Platonic in spite of Kepler's revolutionary advances in the very same era. It was his literal belief in the Copernican system that motivated Galileo's argument and experiments supporting the law of inertia in order to explain to the Aristotelians why, as the earth moves, we are not aware of it.

The modern viewpoint on Newtonian dynamics, informed by general relativity and nonlinear dynamics, abandons the search for global inertial frames. The modern viewpoint emphasizes that freely falling frames are locally inertial. Hence, not all inertial frames can be connected by Galilean transformations. For example, frames in free fall on opposite sides of the earth are both locally inertial, but because both are accelerated toward (or away from) each other they cannot be connected via a Galilean transformation. By abandoning the fiction of global inertial frames and adopting instead the local viewpoint we will not need Newton's global idea of absolute space. The latter requires the unnecessary (and unphysical, because unverifiable) assumption that Cartesian axes can be extended all the way to infinity, an idea that more or less goes back to Descartes, who arrived at the law of inertia independently of Galileo and regarded it as a global law of nature (Galileo's universe was Copernican-Platonic spherical and finite, Descarte's was infinite and unbounded). The noninertial effects identified by Newton as acceleration relative to absolute space are present if we transform from a local frame in free fall to a locally accelerated frame, but the acceleration is relative to a frame that is in free fall relative to the mass distribution that generates the local gravitational field. I will also discuss Mach's criticism of the law of inertia and Newton's second law in part 6 below.

Continuing with the traditional treatment of gravity as a scalar potential, Newton's second law in an inertial frame is given by

$$\frac{d^2 x_i}{dt^2} + \frac{\partial \Phi}{\partial x_i} = 0, \quad (5)$$

and is covariant under Galilean transformations (set $d^2 b_i/dt^2 = 0$ in (8)). The law of inertia in an inertial frame is given by

$$\frac{dv_i}{dt} = 0 \quad (7)$$

and is not merely covariant but is also *invariant* under *Galilean transformations*. This invariance principle (Galilean relativity) describes mathematically the fact that no mechanical experiment can be performed that detects motion at constant velocity relative to a local inertial frame. The mathematical basis for this is that solutions of (7) in two separate inertial frames connected by a Galilean transformation, but using the same initial conditions, are *identical*. Galilean invariance is the basis for the identity of outcome of identically prepared experiments in two separate inertial frames [6].

I should warn the reader that many older books and articles on special and general relativity (and also some newer ones) have (mis-)used the word "invariant" where instead they should have used the word "scalar". For example, statements like "$ds^2 = g_{\mu\nu} dq^\mu dq^\nu$ is invariant" must be replaced by "$ds^2$ is a scalar" under coordinate transformations. The distinction between scalars and invariants is made clear by Hamermesh [7]. Havas [9] discusses Newtonian gravity transforming both as a covariant vector and as an affine connection, but does not distinguish the word "scalar" from the word "invariant". He also states that "... the fundamental equations of Newtonian mechanics are invariant under the Galilei group, those of the special theory of relativity under the transformations of the Lorentz group, and those of the general theory of relativity under all coordinate transformations ("principle of general covariance")", also confusing covariance of vector equations with invariance of differential equations and their solutions.

Covariance of vector equations does not define the physics, which is defined by a principle of relativistic invariance: the law of inertia is covariant with respect to Galilean transformations, but the Galilean relativity principle is reflected by the *invariance* of the law of inertia (7) and its *solutions* under Galilean transformations. Newton's second law is covariant, but not invariant, under the same group of transformations.

Covariance does not carry the weight of a principle that can be imposed externally because it can always be achieved for any equation of motion [10]. Rewriting a law of motion that is correct in a restricted class of frames of reference in covariant fashion does not change the underlying invariance principle that defines the physics [6,11]. As an example, I will show, following Heckmann, how Newton's second law can be made covariant under transformations that include linearly accelerated frames (covariance with respect to transformations to and among rotating frames is also possible [11]). The invariance principle is still Galilean invariance. Cartan [6] showed how Newton's second law can be written in covariant form with respect to transformations to arbitrarily accelerated frames in Newtonian space-time by treating the gravitational force as a "nonintegrable connection". This beautiful geometric picture does not change the fact that Galilean relativity is still the basic invariance principle of classical mechanics. In parts 7 and 8 we see how Cartan brought to light that the only essential difference between Newtonian and Einsteinean mechanics is the replacement of local Galilean invariance by local Lorentz invariance.

**5. Newtonian gravity transforms like a gauge potential**

I start by abandoning the unnecessary assumption that gravity behaves like a scalar potential under arbitrary coordinate changes. Instead, under a transformation from an inertial frame to a linearly accelerated frame,

$x'_i = x_i - b_i(t)$
$v'_i = v_i - \dot{b}_i(t)$
$a'_i = a_i - \ddot{b}_i(t)$  (8)

if we allow the gravitational potential to transform like a gauge potential,

$\Phi'(x') = \Phi(x) + x_i \ddot{b}_i$  (12)

then Newton's second law is covariant with respect to transformations to and among linearly accelerated frames [5],

$$\frac{d^2 x'_i}{dt^2} + \frac{\partial \Phi'}{\partial x'_i} = 0 . \qquad (5c)$$

The covariance of Newton's second law under transformations to accelerated frames does not mean that accelerated frames are equivalent for the purpose of doing experiments in physics [6]. Experiments prepared and performed identically in two differently linearly accelerated frames will necessarily yield different numbers, because the solutions of Newton's laws in these frames are *not invariant* under the transformation (8) whenever $d^2b/dt^2 \neq 0^2$. Whenever $d^2b/dt^2 = 0$ then we retrieve the Galilean transformation and the equivalence in outcome of identically prepared experiments, which is the basis for the human ability to discover laws of nature in the first place [12]. This viewpoint disagrees with Einstein's claims about the physical importance of covariance and the use of arbitrary reference frames for the expression of the laws of physics. The modern viewpoint (emphasized earlier by Fock and Wigner) can be found in the texts by Weinberg [13] and by Misner, Thorne, and Wheeler [14], for example.

If Cartan had thought only about motion in space, rather than in space-time, then he might have arrived at the point of view discussed in the next section.

**6. Newtonian cosmology** *(Alta Via Heckmann)*

The cosmological principle, interpreted globally, would require a matter density that is everywhere constant, with only small fluctuations from its average value. It is well known (see Rindler [15]), but is often ignored in elementary texts and monographs, that a uniform density is impossible in an infinite Newtonian universe. In other words, the cosmological principle can't be realized *globally* in an infinite Newtonian universe, although the same assumption is allowed on a flat 3-torus, where Newtonian mechanics also holds [16].

The first assertion that uniform densities of infinite extent are impossible in Newtonian mechanics is due to Neumann [17]. Milne [18] and McCrea thought that they had discovered a local approach that avoids the global infinity pointed out by Neumann [17,18], but later were proven to have been wrong on that point [19]. Heckmann [5] assumed without checking carefully enough that Milne and McCrea indeed had found a way to avoid the fact that the gravitational force can't be defined globally inside a uniform mass distribution that is infinite in extent. This is not the main point of interest here: Heckmann independently discovered a key result of Cartan's geometric formulation of Newton's laws of motion and gravity, namely, that the principle of equivalence can be built explicitly into Newtonian theory if one assumes that the gravitational potential transforms like a gauge potential under transformations to linearly accelerated frames [5].

I begin with the hydrodynamics description of a pressureless Newtonian dust,

$$\frac{d\vec{v}}{dt} = \frac{\partial \vec{v}}{\partial t} + \vec{v} \cdot \nabla \vec{v} = -\frac{1}{\rho} \nabla \Phi$$

$$\frac{\partial \rho}{\partial t} + \nabla \cdot \rho \vec{v} = 0$$

$$\nabla^2 \Phi = 4\pi\rho \quad , (1)$$

whose characteristic curves, generated by Newton's second law

$$\frac{dt}{1} = \frac{dx_k}{v_k} = \frac{dv_k}{-\frac{\partial \Phi}{\partial x_k}} , \quad (4)$$

and are the trajectories of the dust particles. Solving the quasi-linear partial differential equations (1) is equivalent to solving the nonlinear differential equations (4) with specified initial conditions (and with some prescribed initial density distribution $\rho(x,0)$ at $t=0$ that determines $\Phi(x,0)$)

Under transformations

$$x'_i = x_i - b_i(t)$$
$$v'_i = v_i - \dot{b}_i(t)$$
$$a'_i = a_i - \ddot{b}_i(t) \quad (8)$$

the Abelian gauge transformation rule

$$\Phi'(x') = \Phi(x) + x_i \ddot{b}_i \quad (12)$$

yields covariance of Newton's second law,

$$\frac{d^2 x'_i}{dt^2} + \frac{\partial \Phi'}{\partial x'_i} = 0 . \quad (5c)$$

Note that $b_k$=constant describes translations, $b_k = V_k t$ with $V_k$ = constant describes restricted Galilean transformations, and the frame is linearly accelerated if $d^2 b_k / dt^2 \neq 0$.

A cosmological model is defined by a preferred class of relatively moving coordinate systems. In the preferred frames hypothetical equivalent observers with identical equipment and using equivalent techniques are supposed to be able to observe "the same coarse features of the universe". Clearly, such an assumption will require some yet-to-be specified degree of uniformity of the matter distribution. In Heckmann's model[1] one defines preferred reference frames as those for which the dust velocity field is *invariant* (the analog in general relativity is a maximally-symmetric space, which can be defined via invariance of the metric and mass-energy tensor [13]). When

$$x'_i = x_i - b_i(t) \qquad (8b)$$

we have

$$v_i(x,t) = v'_i(x',t) + \dot{b}_i(t) \qquad (8c)$$

which represents the Newtonian law of combination of velocities. Require in addition that

$$v'_i(x',t) = v_i(x',t) \qquad (13)$$

which means that the velocity field v must be invariant for the preferred class of frames that will be defined by using (13) to determine the functions $b_i(t)$ in (8b).

From (8b), (8c), and (13) we obtain

$$v_i(x,t) = v_i(x - b, t) + \dot{b}_i \qquad (14)$$

so that

$$\frac{\partial v_i(x,t)}{\partial x_k} = \frac{\partial v_i(x - b, t)}{\partial x_k}, \qquad (15)$$

which means that this derivative is independent of x, so that

$$v_i(x,t) = a_{ik}(t)x_k + a_i(t). \qquad (16)$$

I leave it as an exercise for the reader to show that $a_i=0$ is required, yielding

$$v_i(x,t) = a_{ik}(t)x_k \qquad (17)$$

which is a generally anisotropic Hubble law of expansion or contraction of the matter distribution.

To go further we need the equation of continuity, which we can more conveniently impose in the particle mechanics form [6]

$$\prod_{i=1}^{3} dx_i(t) = J(t) \prod_{i=1}^{3} dx_{io}. \qquad (18)$$

Here, $x_i(t)$ is the position of a dust particle at time t, $x_{io} = x_i(0)$ is the initial condition at $t = 0$, and $J(t)$ is the Jacobian of the one parameter transformation from the variables $x_{io}$ to the variables $x_i(t)$. It is well-known that

$$\dot{J} = \nabla \cdot \vec{v} J \qquad (19)$$

where v is the velocity field of the dust particles in the six dimensional phase space. Combining this with the Hubble law (17) yields

$$\frac{\dot{J}}{J} = \nabla \cdot \vec{v} = \text{Tr } a(t) \qquad (20)$$

where a(t) is the 3x3 matrix in the Hubble law (17). Hydrodynamicists might say that we are working in the Lagrangian rather than Eulerian picture, but we are simply studying the characteristic equations (4) by standard methods of nonlinear dynamics. The Lagrangian picture of hydrodynamics amounts to integrating Newton's equations (5) backward in time while using the fact that the initial conditions $x_{io}$ are trivially conserved along characteristics. If the solutions of (5c) do not have spontaneous singularities (singularities at real, finite times) then we have a flow in a six dimensional (x,v) phase space, where the trajectories can be thought of as streamlines, and the transformation from the $x_{io}$ to the $x_i(t)$ can be regarded as a one-parameter coordinate transformation with Jacobian J(t).

If the initial conditions and dynamics permit the definition of a once-differentiable matter density ρ then

$$\frac{d\rho}{dt} = \dot{\rho} + \vec{v}\cdot\nabla\rho = -\rho\nabla\cdot\vec{v} = -\rho\frac{\dot{J}}{J} \qquad (21)$$

along dust particle trajectories so that

$$\rho = \rho_o/J(t) \qquad (22)$$

along a characteristic curve, where $\rho_o$ is independent of t. In chaotic nonlinear dynamics initial conditions may not permit the definition of a smooth pointwise density but one can always work with coarsegrained pictures where, at least initially, densities are piecewise constant.

The gravitational potential obeys

$$-\frac{\partial \Phi}{\partial x_i} = \dot{v}_i + v_k \frac{\partial v_i}{\partial x_k}$$
$$= -\dot{a}_{ik}x_k - a_{ik}a_{kl}x_l \qquad (23)$$

and with a generally anisotropic Hubble law $v_i = a_{ik}x_k$ we obtain

$$\frac{\partial^2 \Phi}{\partial x_i \partial x_k} = \frac{\partial^2 \Phi}{\partial x_k \partial x_i}, \qquad (24)$$

This condition guarantees that $\Phi$ exists and yields $da_{ik}/dt = da_{ki}/dt$, which means that a is a symmetric matrix,

$$a_{ik}(t) = a_{ki}(t). \qquad (25)$$

Therefore, the gravitational potential corresponding to the anisotropic Hubble motion (17) is given by

$$\Phi(x,t) = -\frac{1}{2}(\dot{a}_{ik}(t) + a_{il}(t)a_{kl}(t))x_i x_l. \qquad (26)$$

It is easy to show that the potential is invariant under the transformations (8):

$\Phi(x',t) = \Phi'(x',t)$.

We can now see the impossibility of a static Newtonian universe, where no evolution could occur. With

$$\nabla \cdot \vec{v} = \text{Tr } a(t) = -\frac{1}{\rho}\frac{d\rho}{dt} \qquad (27)$$

and

$$\nabla^2 \Phi = -\text{Tr } \dot{a}(t) - \sum_{i,k} a(t)_{ik}^2 = 4\pi G\rho \qquad (28)$$

we find that

$$\frac{d}{dt}\frac{1}{\rho}\frac{d\rho}{dt} = \sum_{i,k} a(t)_{ik}^2 + 4\pi G\rho \geq 0 \qquad (29)$$

Therefore $d\rho/dt = 0$ is impossible and $\rho(t)$ cannot be a constant. Is is still possible that the density $\rho$ is *spatially* constant for fixed times (is the cosmological principle realizable in this model)? I defer the analysis until the end of this section, assuming temporarily without proof that a uniform and isotropic Newtonian universe is possible at any given time t.

Consider next only isotropic Hubble motions. Starting with

$$v_i(x,t) = a_{ik}(t)x_k \qquad (17)$$

and

$$\nabla \cdot \vec{v} = \operatorname{Tr} a(t) = -\frac{\dot{J}}{J} \quad (30)$$

and then imposing the isotropy condition

$$a_{ik}(t) = \delta_{ik} \frac{\dot{J}}{3J}, \quad (31)$$

we find that can write

$$v_i(x,t) = \frac{\dot{J}}{3J} x_i \quad (32)$$

where it follows that the Hubble 'constant' is given by $H(t) = (dJ/dt)/3J$. Since the dust particle trajectories are generated by

$$\dot{x}_i = \frac{\dot{J}}{3J} x_i \quad (33)$$

and one integration yields the time evolution rule

$$x_i(t) = (J(t))^{1/3} x_{io}. \quad (34)$$

Writing $dR/dt = (dJ/dt)/3J$ yields Hubble's law in the form

$$v_i(x,t) = \frac{\dot{R}}{R} x_i \quad (33b)$$

so that the dust particle trajectories can be written as

$x_i(t) = R(t) x_{io}$. (34b)

The time evolution of the dust particles is just a rescaling of the initial conditions $x_{io}$ where the scale factor is $R(t) = (J(t))^{1/3}$, and $J(t)$ is the Jacobian of the time evolution transformation. We can also write

$$\rho(t) = \frac{\rho_o}{R(t)^3}. \quad (35)$$

Next, I want to solve for the parameters $b_i(t)$ in order to see what the preferred class of equivalent reference frames looks like. Combining the Newtonian law of addition of velocities (8) with the generalized Hubble motion (17) and the invariance condition (13) that defines our cosmology we obtain

$$\begin{aligned} v_i(x',t) &= a_{ik}(t)x_k - \dot{b}_i(t) \\ &= a_{ik}(t)(x'_k + b_k(t)) - \dot{b}_i(t) \\ &= a_{ik}(t)x'_k \end{aligned} \quad (36)$$

This yields the differential equations

$$\dot{b}_i = a_{ik}b_k \quad (37)$$

that define the preferred frames.

In the isotropic model, $a_{ik} = \delta_{ik} (dJ/dt)3J = \delta_{ik} (dR/dt)R$, we have

$$\dot{b}_i = \frac{\dot{R}}{R} b_i \qquad (38)$$

which yields

$$b_i(t) = cR(t) \qquad (39)$$

with c a constant. Therefore the preferred frames are defined by the one parameter coordinate transformations (c is the parameter)

$$x'_k = x_k - cR(t), \qquad (40)$$

where $R(t) = (J(t))^{1/3}$. Translational invariance would require that we find an additive constant when we solve (the Friedmann equation) for R(t). Globally seen, the preferred frames are not inertial frames (we will see later that the form of R(t) rules out global Galilean invariance). However, because each frame is in free fall in the gravitational field of the other N-1 dust particles, these frames are locally inertial.

We now obtain the equation of motion for the expanding universe. Combining

$$- \mathrm{Tr}\, \dot{a}(t) = \sum_{i,k} a(t)_{ik}^2 + 4\pi G\rho \qquad (41)$$

with the local isotropy assumption $a_{ik} = \delta_{ik}(dR/dt)/R$ with $\rho = \rho_o/R^3$ we find

$$-3\frac{d}{dt}\frac{\dot{R}}{3R} = \left(\frac{\dot{R}}{R}\right)^2 + \frac{4\pi G\rho_o}{R^3} \qquad (42)$$

which we can rewrite as Newton's second law for free fall in the field of a point singularity

$$\ddot{R} = -\frac{4\pi G\rho_o}{R^2} \quad (43)$$

at $R = 0$ (since we are doing classical mechanics, this result should give us a hint that this universe has a center). Newton's equation of motion can be integrated once to yield the Friedmann equation [5]

$$\frac{\dot{R}^2}{2} - \frac{4\pi G\rho_o}{R} = \varepsilon = \text{constant} \quad (44)$$

Note that R is only a scale factor so that one should not interpret $\rho_o$ as the mass of the universe. With $\varepsilon = 0$ we have expansion with $dR/dt=0$ at $R=\infty$. $\varepsilon>0$ yields expansion with finite expansion rate at infinity, and when $\varepsilon<0$ there is expansion to a finite value $R_m = 4\pi GM/(-\varepsilon)$ followed by collapse to $R=0$ in finite time (spontaneous singularity of equation (43)).

The cosmological principle appears to hold locally: the density is spatially uniform except near the boundary of the mass distribution, which is ignored in this analysis (boundary conditions on $\Phi$ at large $x_i$ were never discussed), and except near the point of gravitational collapse (the singularity of the mass distribution). I will explain below why the mass distribution is necessarily finite in extent, why the uniform density must be cut off after a finite distance, and where the finite-sized boundary of the mass distribution was swept under the rug in Heckmann's treatment. Locally, inside the mass distribution and far from the edges of the finite universe (imbedded in an infinite Euclidean space), the density is spatially uniform, the universe is locally homogeneous and isotropic, so that the cosmological principle holds locally (Milne, McCrea, and Heckmann had believed that the cosmological principle would hold globally in this model).

A Newtonian universe where the cosmological principle holds globally cannot have a singularity in time for the density because a spontaneous singularity of (43) defines the center of the universe. I stress that this finite time singularity of Newton's law is not the same geometrically/qualitatively as the collapse of the entire space-time manifold in general relativity, although the Friedmann equation (43) describes both cases quantitatively.

Where is the Neumann-Seeliger infinity hidden in the previous analysis if we would assume, as did Heckmann (and later Bondi [20], who repeated the simpler Milne-McCrea analysis), that we can take $\rho(t)$ to be spatially constant all the way to infinity?

I will show first that the universe defined by our model has a center. A hypothetical observer who's near neither the edge nor the center of the mass distribution won't know in advance where the center is located before the collapse occurs, unless he's carried out calculations that go beyond the scope of these lectures. In other words, a preferred observer sees the contraction that occurs for $\varepsilon < 0$ but, because of invariance of the velocity field, all observers see the same contraction so that none of these observers can say in advance where the collapse is going to take place. This center differs both qualitatively and quantitatively from the anthropic center of the universe that Plato and Aristotle imagined, and also from Copernicus' neo-Platonic godly center.

To see why there is a center consider the case where $\varepsilon<0$:

1. Gravitational collapse occurs in finite time (R(t) vanishes in finite time).
2. $x_i(t) - x_j(t) = R(t)(x_{io} - x_{jo})$ vanishes as the scale factor R vanishes; all particle displacements collapse as R vanishes.
3. $x_i(t) = R(t)x_{io}$ vanishes as R vanishes, so that each and every dust particle approaches the origin of its coordinate system as the collapse proceeds.
4. $x'_k = x_k - R(t)$ approaches $x_k$ as R vanishes, so that all coordinate origins coincide at the time of gravitational collapse.

*The entire mass distribution disappears into a spontaneous singularity of the nonlinear differential equation (43) in finite time.* In other words, this model of the universe has a center.

Furthermore, the uniform mass distribution cannot be infinite in extent and must have a boundary. Where was the violation of this condition hidden in the analysis? By introducing the idea of the density and then assuming that $\rho(t)J(t) = \rho_o$ is constant we *implicitly* restricted our analysis to local internal regions far from any boundary, where the mass density (if it exists) may look nearly uniform. If we would assume that the same condition could hold all the way to infinity then we would not be able to define either $\Phi(x,0)$ or the gravitational force. In other words, the Neumann-Seeliger infinity was hidden in the initial condition $\rho_o$. Milne, McCrea and Heckmann did not face the infinity (that infinity informs us that a uniform mass distribution cannot be spatially infinite) because they ignored the boundary conditions on the potential Φ (Milne had offered his model as an example of how pure mathematics can be used to short-circuit the need for physics [17]).

In a global analysis the velocity field and potential cannot be invariant under the transformations (8) that define the preferred observers locally. The expansion/contraction does not look the same for observers near the edge as for observers far from the edges. An observer with a global viewpoint (a space traveler outside the mass distribution, e.g.) could know where the collapse is going to occur, while an observer well within the interior (who hasn't done the necessary calculations) would be ignorant of the fact that the universe has a center. The size of the universe can be taken as proportional to the scale factor R(t). For a spherical universe the center of the universe lies at the sphere's center. The cosmological principle doesn't hold globally but holds locally well within the interior, at times that are not too near to the collapse time.

**7. Relativism or relativistic invariance?**

"An influence of the local inertial frame on the stars is not acceptable, and hence it must be assumed that *the local inertial frame is determined by some average of the motion of the distant astronomical objects.* This statement is known as Mach's principle."

<div style="text-align: right">H. Bondi, in <u>Cosmology</u> [20]</div>

It is difficult to discuss Mach's principle. There is still no convincing realization of Mach's Principle. The number of different statements of Mach's principle may be on the order of magnitude of the number of different writers on the subject. Interesting attempts to realize Mach's principle in the context of classical mechanics can be found in references [21,22,23,24,25].

Mach himself offered no principle but criticized both the law of inertia and Newton's second law. Mach [26] wrote from a standpoint of relativism in philosophy and physics (Descartes and Huyghens were earlier advocates of relativism[2]). Roughly speaking, Mach asserted that only relative positions, velocities, etc. between masses should enter into laws of motion, and that mechanics should not be local: the entire universe should be considered in defining inertia. This is a holistic point of view. I will argue below that local laws of motion are enough, that holism and relativism are unnecessary.

Mach's call for relativism is not the same as a principle of relativistic invariance (Galilean or Lorentzian), and (upon closer inspection) amounts to a criticism of principles of relativistic invariance because those principles are based on translational invariance and invariance with respect to transformations to uniformly moving frames relative to a local inertial frame that replaces Newton's idea of a fixed, global inertial frame. Mach criticized the law of inertia [26] (which is essentially the same as criticizing Galilean, and also Lorentzian, invariance) as not having been derived from an equation of relative motion of interacting bodies. Einstein was stimulated by the criticism of Newtonian and Galilean ideas found in Mach's historico-philosophical book [26], as were many other physicists, psychologists, and philosophers [29], but Einstein did not succeed in incorporating Mach's demand for relativism into physics, or even in clarifying what we should understand as Mach's principle. I will explain next how Mach's relativismic (as opposed to relativistic) criticism of Galileo and Newton can be short-circuited. It is not an accident, or a mere misfortune of old age (as some authors have lamented), that Mach did not see his ideas of relativism reflected in Einstein's work on either special relativity or gravity.

---

[2] Barbour [28] argues that Huyghen's belief in Cartesian mechanism prevented his getting credit for both Newton's second law and the law of gravity. Newton, in contrast, hated Cartesian relativism and was driven to dispute it [29]. Newton understood the difference between (Galilean) relativity and relativism. Descartes adopted a position of relativism in an attempt to avoid the Index and the inquisition. By asserting that the difference between taking the sun to be fixed or the earth to be fixed (choice of coordinate system, in our language) is merely a matter of convenience, he hoped to avoid being charged with disputing the notion that the earth may stand still. Following Kepler, (whom Galileo completely ignored) Newton showed that this is not merely a matter of convenience, that the sun constitutes a better approximation to an inertial frame than does the earth for the calculation of planetary motions. One can calculate planetary motions from the earth's frame if one wishes, but the calculations (obtained via transformation from an inertial frame) will be much more complicated [30].

Mach wanted to define forces and relative motions first, and then derive the law of inertia from a reformulated second law of Newton (Einstein did the opposite--see part 8). He was misled by the pre-Einsteinean assumption that inertial frames are *globally* possible. We now expect that they are not, and we can reinterpret Newtonian physics with the benefit of both Einstein's contributions and insight provided by modern nonlinear dynamics. This is done, following Cartan, in the next section. The following point of view eliminates Machian-style objections to the appearance of inertial terms interpreted as accelerations relative to absolute space, where absolute space is usually regarded as some undefined collection of global inertial frames, or as one global frame of reference fixed in the (only approximately) "fixed stars". In place of this global picture we can instead adopt the following position, which relies upon and emphasizes the typical nonintegrability of local laws of motion.

A neutral body subject to no other net force is always locally in free fall in the net gravitational field of the rest of the matter in the universe. A body in free fall defines a local inertial frame. *There need be no other inertial frames in the universe.* Noninertial effects then occur relative to a local frame that is accelerated relative to a local inertial frame, and therefore occur indirectly relative to the masses that determine the *local* gravitational field over laboratory or observational times (the earth falls freely about the sun, the moon falls freely about the earth in the field of both the earth and sun, etc.). A local frame is approximately inertial only for a finite time. In that frame the law of inertia holds,

$$\frac{dv_i}{dt} = 0, \qquad (7)$$

where we can [6] and should think of the Cartesian axes $x_i$,

$$x_i(t) = v_i t + x_{io}, \qquad (45)$$

as *generated* by the motion of three free tracer-particles moving rectilinearly with constant speeds $v_i$. *In other words, the law of inertia (7) generates the local inertial frame, whose axes are defined by (45).* Mathematically, (7) is globally integrable in Newtonian mechanics but we ask next whether the applicability of this formal mathematical condition might be preempted by the physics of tracer particles. Given finite velocities $v_i$, the spatial extent of the Cartesian axes (45) would be limited by the finite time $t_{max}$ over which the local frame is approximately inertial, which is determined by the neglected gradients of the local gravitational field. Also, for any finite velocity v the differential equation (7) that generates the axes (45) locally will fail to hold at large times $t \gg t_{max}$ because of the (neglected) field gradients, so that (7) would have to be replaced globally by (5). Even in classical mechanics there is no way to generate infinitely-long straight lines via particle motions when it is taken into account that matter is distributed throughout the universe. However, in classical mechanics the speeds $v_i$ can be as large as you like so that, in principle, the axes of the inertial frame can be imagined to extend spatially as far as you like (corresponding mathematically to the fact that (7) is globally integrable in a flat space) in spite of the fact that the frame is approximately inertial only over a finite time $t_{max}$. Summarizing, in classical mechanics a local inertial frame (45), with axes fixed in a freely falling body, can only be imagined to extend in all three directions to infinity by using the unphysical artifice of infinite tracer-particle velocities, but still that local frame is approximately inertial only for a finite time.

Special relativity limits all speeds to no more than the speed of light. Therefore, local inertial frames can be realized in a gedanken experiment by tracer particles (including photons) only for finite times and are also finite in extent. Cartesian inertial frames extending to infinity cannot be constructed in a gedanken experiment that takes into account both the speed of light and the nonemptiness of the universe (gravity can be eliminated locally, but not globally in cosmology). As Havas [19] has noted, general relativity has two distinct mathematical limits where global inertial frames are allowed mathematically (the resulting three-space is Euclidean, (7) is globally integrable, and so (45) holds mathematically, if not physically, for $-\infty \leq t \leq \infty$): (1) neglect gravity but keep Lorentz invariance (special relativity in an empty

universe[3]), or (2) let the speed of light go to infinity but keep gravity (local Galilean invariance with Newtonian gravity).

"... - the independence of the laws of nature from the state of motion in which it is observed, so long as it is uniform - is not obvious to the unpreoccupied mind. One of its consequences is that the laws of nature determine not the velocity but the acceleration of a body..."

E. P. Wigner [12]

In contrast, Mach refused to regard the law of inertia as an independent law of motion. He regarded it instead as already defined by the statement that acceleration vanishes when net force vanishes. This is a superficial viewpoint, but a viewpoint that one can easily adopt by default through failing adequately to appreciate that Galilean invariance is the foundation of Newton's second law [6,12].

"... [the Mach Principle] ... is the tendency to derive the meaning ... from the whole ... ."

Otto Neurath [29]

Mach's vague, holistic idea of obtaining the law of motion of interacting bodies first and then deriving the law of inertia from it puts the cart before the horse. However philosophically appealing and invulnerable Mach's criticism might have seemed earlier, we can now see it's weaknesses: Wigner [12] has pointed out the necessity of local invariance principles (like Galilean relativity) as the basis for our ability to discover of laws of nature in the first place. Galileo's two local laws (the law of inertia and the principle of equivalence) were necessary before the second law could be formulated by either Huyghens (for uniform circular motion only) or Newton (most generally). Mach's objections to Newtonian mechanics were fueled by the mistaken (and historically-understandable) assumption that inertial frames, infinite in extent and defined by "the fixed stars", instead of local frames defined by freely falling bodies, could exist for arbitrarily long times.

---

[3] One can solve the Kepler problem in special relativity [6], but that is not the point here.

Summarizing, physics is described locally by differential equations. The differential equations of mechanics are universally applicable, locally, so long as we can find approximate inertial frames. Inertial frames are fixed in locally freely-falling bodies (gravitational interactions are always present in the real universe), and are therefore approximately inertial for limited times only. Whether or not global laws can be deduced from the local laws of motion is a question of integrability. From different geometric perspectives both modern nonlinear dynamics and differential geometry inform us that most systems of differential equations are nonintegrable in one sense or another, which means that the calculation of correct predictions for very long times, and over very long distances cannot be taken for granted. The realization that correct global results are very, very hard to determine leads one to the viewpoint that cosmology, in the end, may not produce much more than locally correct results.

I have argued elsewhere [31] that an analog of the law of inertia would be necessary for economics, sociology, and psychology before there could be any hope to retrieve those fields from the mathematical lawlessness that is largely their present content. The existence of such an analog is unlikely because people, unlike billiard balls and planets, can either cause or avoid collisions simply by changing their minds either systematically or arbitrarily. The ability to change one's mind is a prime example of mathematical lawlessness: a dynamical system (like the Newtonian three body problem) cannot change it's orbits arbitrarily, and cannot learn from the past. The absence of fixed mathematical law in brain-driven "motions" is the reason why *artificial* "law" is legislated (the motions of a deterministic or probabilistic dynamical system cannot be legislated). To argue that brain-driven motions may be more like a neural network that learns, than like a Newtonian dynamical system, is the same as admitting that brain-driven behavior is effectively mathematically lawless (such a system's dynamics must change unpredictably as it learns unexpectedly).

Mach's thinking fed directly into and reinforced philosophical relativism [28], whose extreme wing (found mainly in literary criticism [32], cultural studies, anthropology, and sociology) now attacks physics as just another arbitrary activity like sociology, where there are no universal laws and where a "text" is presumed without proof to have no more meaning than a collection of abstract symbols on a printed page [33]. Interpretations of "texts" are regarded merely as arbitrary "representations" in the imprecise and undefined jargon of postmodernism.

"... we have to consider science as a human enterprise by which man tries to adapt himself to the external world. Then a "pragmatic" criterion means ... the introduction of psychological and sociological considerations into every science, even into physics and chemistry. ....the sociology of science, the consideration of science as a human enterprise, has to be connected in a very tight way with every consideration which one may call logical or semantical."

Philipp Frank [29]

Relativismic socio-literary criticism may describe contemporary literary theory and the confusion within the socio-economic sciences, but it fails to shed light on scientific fields like physics, chemistry, and genetics that are grounded in empirically-established local invariance principles, and in empirically-verified universally-applicable local laws of motion of interacting bodies that are based on those invariance principles.

**8. Newtonian gravity is a nonintegrable connection in space-time**

Cartan [4,6,14,34], with hindsight informed by general relativity, noticed that we can rewrite Newton's second law for motion of a body of mass m in a net gravitational potential $\Phi$

$$\frac{d^2 x_i}{dt^2} + \frac{\partial \Phi}{\partial x_i} = 0 \qquad (5)$$

in the following way:

$$\frac{d^2 x^k}{d\lambda^2} + \frac{\partial^2 \Phi}{\partial x^{k2}}\left(\frac{dt}{d\lambda}\right)^2 = 0, \quad \frac{d^2 t}{d\lambda^2} = 0 \qquad (46)$$

The time t is linear in $\lambda$ and is is affine. We can always choose t=$\lambda$. Go next to space-time coordinates $x^\mu = (t, x^1, x^2, x^3)$. Phoenician letters denote spatial components (1,2,3) of coordinates, vectors, tensors, and connections. Note that we can write

$$\ddot{x}^\mu + \Gamma^\mu_{\nu\lambda}\dot{x}^\nu\dot{x}^\lambda = 0 \qquad (47)$$

where

$$\Gamma^i_{oo} = \frac{\partial \Phi}{\partial x^i}, \; \Gamma^a_{bc} = 0 \qquad (48)$$

Newtonian space-time is not flat because the Riemann curvature tensor is given by

$$R^i_{ojo} = \frac{\partial^2 \Phi}{\partial x^j \partial x^i}, \; R^a_{bcd} = 0 \qquad (49)$$

Three-space is flat with Cartesian coordinates $x^i$, and time is absolute (simultaneous spatially-separated events are allowed), but the four coordinates $x^\mu$ are not globally Cartesian on four-dimensional space-time manifold because of curvature in the three $(t, x^i)$ sub-manifolds. Newtonian gravity is therefore not a covariant vector, but is instead a nonintegrable connection $\Gamma$ in space-time. Kepler's orbits, as described by Newton, are therefore geodesics in curved space-time. Note that Heckmann's treatment of the gravitational potential $\Phi$ as a gauge potential, $\Phi(x',t) = \Phi(x,t) + x_k(d^2b^k/dt^2)$, under transformations $x'^k = x^k - b^k(t)$ from inertial to linearly accelerated frames, and correspondingly the principle of equivalence, are automatically built into Cartan's interpretation of Newton's laws (equation (5) holds whether the frame is locally inertial, as is the earth in it's motion about the sun, or is linearly accelerated, like a train leaving sentralstasjon in Oslo).

Continuing, we find that the Ricci tensor is given by

$$R_{oo} = \nabla^2 \Phi = 4\pi\rho, \quad R_{ab} = 0. \qquad (50)$$

Newtonian space-time is not a metric space: a nondegenerate metric g cannot be defined that is consistent with the covariant derivative [14] (Heckmann may not have been aware of this restriction).

This formulation can be extended to include transformations to rotating frames as well [6,14]. In other words, Newtons' second law is covariant with respect to transformations to arbitrarily accelerated frames in space-time, where the local relativity principle is Galilean invariance. General covariance plays no physical role here or in general relativity. Arbitrarily accelerated frames are certainly not equivalent for the performance of identically-prepared mechanical experiments. Local inertial frames are clearly preferred, for otherwise there is no identity of outcomes for experiments with identical preparation (solutions of (7) in two different frames connected by a Galilean transformation, but using identical initial conditions, are identical).

A recent paper [35] on knot theory implicitly presumes Cartan's interpretation of Newtonian mechanics/gravity (geodesics in space-time due to the equivalence principle), but without reference or explanation.

Had Einstein been able to sidestep Mach and global inertial frames ("absolute space") he would, in principle, have been able to arrive more directly at the formulation of general relativity. Or, as Thomas Buchert [36] has put it, "If Lagrange (could) have had a beer with Newton, then they could have derived the Einstein-Cartan theory" (on a nice Biergarten tablecloth, perhaps).

By starting with Cartan's description of Newtonian mechanics, and then replacing local Galilean invariance by Lorentz invariance, one arrives at the doorstep of general relativity, which is not a new or global relativity principle at all but is instead a locally Lorentz invariant description of gravity based upon the principle of equivalence. Cartan showed how to build the principle of equivalence into Newtonian mechanics explicitly. The equivalence principle was built into Newtonian mechanics independently and later by the cosmologist Heckmann. Heckmann seems to have been unaware of Cartan's work, which shows that Newtonian motion in a gravitational field can be understood as geodesic motion in space-time, where space is Euclidean and time is absolute. There are now attempts to use Cartan's interpretation of Newtonian dynamics as the framework for formulating and doing averaging and perturbation theory.

## 9. Einstein's theory of gravity

To arrive at general relativity we could simply start with Cartan's description of Newtonian mechanics and replace local Galilean invariance by local Lorentz invariance [37]. Instead, I will follow an alternative path that uses only local Lorentz invariance, the local law of inertia (7), and the principle of equivalence (see also [13] or [14]). In other words, we start locally, making no global assumptions. Global results, if they exist at all, must follow from global integrability of the resulting local law of motion.

The law of inertia in any local Lorentz frame is given by

$$\frac{d^2 x^\mu}{d\tau^2} = 0 \qquad (7b)$$

where $\tau$ is the proper time.

The law of inertia is globally integrable to yield Cartesian coordinates

$$x_\mu(\tau) = v_\mu \tau + x_{\mu o} \qquad (51)$$

(the $v_\mu$ are constants) if and only if the Riemann curvature tensor vanishes everywhere (although no one need be concerned mathematically about distances greater than the speed of light times the age of the universe). Global integrability of (7b), with no points of the manifold excluded, is the formal mathematical (but not necessarily physical) condition to extend local Cartesian axes defined by (51) all the way to infinity (the space-time manifold is then globally flat). This is equivalent to the assumption that the Lorentz metric $\eta_{\mu\nu}$ is globally diagonal and constant, and so is given by $\eta = (-1,1,1,1)$.

Instead of assuming that space-time is globally flat, begin by transforming the locally Lorentz invariant law of inertia (7b) to any other local coordinate system $q^\mu = f^\mu(x)$, $x^\nu = h^\nu(q)$,

$$0 = \frac{d^2x^\mu}{d\tau^2} = \frac{d^2q^\nu}{d\tau^2}\frac{\partial h^\mu}{\partial q^\nu} + \frac{\partial^2 h^\mu}{\partial q^\alpha \partial q^\nu}\dot{q}^\alpha \dot{q}^\nu. \quad (52)$$

So far, we have only assumed that the space-time manifold is locally Lorentzian (therefore locally flat).

Using

$$\frac{\partial f^\mu}{\partial x^\nu}\frac{\partial h^\nu}{\partial q^\kappa} = \delta_{\mu\kappa} \quad (53)$$

we obtain the law of inertia

$$\frac{d^2q^\mu}{d\tau^2} + \Gamma^\mu_{\alpha\beta}\dot{q}^\alpha \dot{q}^\beta = 0 \quad (50b)$$

in any other coordinate system. The affine connections are given by

$$\Gamma^\mu_{\alpha\beta} = \frac{\partial h^\mu}{\partial x^\kappa}\frac{\partial f^\kappa}{\partial q^\alpha \partial q^\beta} \quad (54)$$

and include all noninertial effects like linear and angular accelerations relative to freely falling frames, and also (according to Cartan [34] and Einstein [38]) global gravitational effects as well. The principle of equivalence was used implicitly because we cannot distinguish gravity locally from a linear acceleration.

The q's are called generalized coordinates. Synonyms for generalized coordinates are holonomic coordinates, or just coordinates [6,14][4]. Generally, the coordinates $q^\mu$ also do not exist globally (the differential equations that generate them are only locally integrable) unless they reflect the symmetry of the manifold, like spherical coordinates in the Schwarzschild solution of general relativity. Again, the condition for global parallelism, represented by the extension of (51) to $\tau = \pm\infty$, is that the Riemann tensor

$$R^\alpha_{\eta\gamma\beta} = \frac{\partial \Gamma^\alpha_{\beta\eta}}{\partial q^\alpha} - \frac{\partial \Gamma^\alpha_{\eta\gamma}}{\partial q^\beta} + \Gamma^\alpha_{\tau\gamma}\Gamma^\alpha_{\tau\gamma} - \Gamma^\alpha_{\tau\beta}\Gamma^\tau_{\gamma\eta} \qquad (55)$$

vanishes everywhere. This represents the essential physics of the transition from special relativity and Newtonian mechanics to general relativity.

The next step would be to generalize the Cartan-Newton equation (50) for the gravitational potential to include a locally-conserved mass-energy tensor on the right-hand side. Gravity is here described by a nonintegrable connection $\Gamma$ in a locally Lorentzian space-time, and therefore is determined by the ten gravitational potentials $g_{\mu\nu}$ (in the absence of rotation, and using holonomic coordinates, $\Gamma$ is symmetric). Starting with a Cartesian frame $x^\mu$ in free fall and the local Lorentz metric $\eta = (-1,1,1,1)$, where $ds^2 = \eta_{\mu\nu}dx^\mu dx^\nu$, the transformation to any other local coordinate system yields

$$ds^2 = \eta_{\mu\nu}dx^\mu dx^\nu = g_{\alpha\beta}dq^\alpha dq^\beta \qquad (56)$$

---

[4] A crude dictionary relating turn-of-the-century phrases to modern terminolgy in differential geometry reads in part as follows: 'holonomic coordinates' [6] represent a 'coordinate basis' [14] while 'nonintegrable velocities' [6] correspond to a 'noncoordinate basis' [14].

where the metric is given at least locally by

$$g_{\alpha\beta} = \eta_{\mu\nu} \frac{\partial h^{\mu}}{\partial q^{\alpha}} \frac{\partial h^{\nu}}{\partial q^{\beta}}. \qquad (57)$$

The principle of equivalence combined with local Lorentz invariance, not general covariance, is the basis for general relativity, which is not a global relativity principle at all (and is also not a theory based on relativism in Mach's sense) but is a geometric theory of gravity. The principle of equivalence as the physical basis for the geometric theory of gravity was first emphasized by Einstein [38].

In another paper [39] I will discuss the observed distribution of the galaxies and ask whether there is any evidence to support either the cosmological principle or a hierarchical universe, e.g, a fractal universe.

**Acknowledgement**

These notes reflect in part lectures presented at the University of Oslo in the spring of 1997 during my sabbatical, during which time I was a guest of both the Institute for Theoretical Physics at the University and the Physics Department of The Institute for Energy Technology. I am grateful to Jan Frøyland and Arne Skjeltorp for guestfriendship, and to Finn Ravndal and Øyvind Grøn for persistently stimulating questions and criticism during my lectures. Earlier in my sabbatical, from September 1996, through February 1997, I was guest professor in Lehrstuhl Wagner at Ludwig Maximillian's Universität in München, where Herbert Wagner and Thomas Buchert introduced me to Heckmann's papers, which are not available in English. Part 7 can be seen more or less as an edited and extended version of an informal translation of the Newtonian part of Heckmann's *Theorie der Kosmologie*. I am also grateful to Julian Barbour for friendly correspondence and criticism of my discussion of Mach's principle and relativism. This work was supported financially by The Research Council of Norway, the Institute for Energy Technology (at Kjeller, Norway), and the Departments of Physics at the Universities of Oslo (UIO), München (LMU) and Houston (UH).